# Pair-Breaking, Pseudogap, and Superconducting $T_c$ of Hole Doped Cuprates: Interrelations and Implications

S. H. Naqib*, R. S. Islam, Ihtisham Qabid

*Department of Physics, University of Rajshahi, Rajshahi-6205, Bangladesh*

**Abstract** Irrespective of the class they belong to, all the hole doped high-$T_c$ cuprate superconductors show an anti-correlation between the superconducting transition temperature and the characteristic pseudogap energy in the underdoped region. The doping dependent pseudogap in the quasiparticle spectral density is believed to remove low-energy electronic states and thereby reduce the superconducting condensate. Impurities within the $CuO_2$ plane, on the other hand, break Cooper pairs in the unitarity limit and diminish superfluid density. Both pseudogap in pure cuprates and impurity scattering in disordered cuprates reduces $T_c$ very effectively. In this study we have compared and contrasted the mechanisms of $T_c$ degradation due to pseudogap and impurity scattering in hole doped cuprates. We have suggested a framework where both these factors can be treated on somewhat equal footing. Beside impurity and pseudogap dependent superconducting transition temperature, the proposed scenario has been employed to investigate the disorder and hole content dependent isotope exponent in high-$T_c$ cuprates in this work.

## 1 Introduction

The emergence of superconductivity in hole doped cuprates has baffled the condensed matter physics community still after over three decades since its discovery [1 – 3]. The Mott insulating parent phase with antiferromagnetic (AFM) order becomes metallic when around 0.05 holes ($p$) are added to the $CuO_2$ planes. AFM long range order disappears and a superconducting (SC) ground state is formed upon hole doping. The SC transition temperature, $T_c$, rises rapidly with increasing hole content, attains a maximum ($T_{cmax}$) at the optimum doping ($p \sim 0.16$), starts decreasing with further (over) doping [4], and vanishes eventually in the deeply overdoped region ($p \sim 0.27$). The $T_c(p)$ of different families of hole doped cuprates exhibits a nearly parabolic dependence [5 – 7].

*Corresponding author. Email: salehnaqib@yahoo.com

It is thought that various properties in the normal and SC states of these strongly correlated electronic systems are non-Fermi liquid like [8 – 10] over a wide range of hole contents. The electronic temperature – hole content (*T-p*) phase diagram shows interplay among a number of electronic ground states (*e.g.*, AFM Neel order, spin glass state, charge and spin density orders, pseudogap (PG) region, SC phase, and a more Fermi-liquid-like metallic state as one increases the hole content) [9 - 11]. All these pose a serious theoretical hurdle to formulate a coherent scheme describing these varied facets as doping levels are changed. The conventional BCS formalism [12], developed for weakly interacting metallic compounds, is unable to describe various features shown at and below the SC transition temperature in hole doped cuprates.

Almost all of the normal and SC state properties are affected significantly by the magnitude of a PG in the low-energy quasiparticle (QP) spectral density of states [6 – 9, 13 – 15]. PG dominates the the superfluid density, pair condensation energy, electronic entropy, isotope exponent, and critical current density [13 – 16] in the SC state. All normal state transport and magnetic properties are also dominated by the PG [17, 18]. It is an experimentally well established fact that the PG removes low-energy QP spectral weight which otherwise would have been available for pairing condensate [17, 18]. As a result, the superfluid density is diminished. At the same time, experimentally it is also found that, the SC transition temperature varies almost linearly with the superfluid density in the underdoped (UD) region of the hole doped cuprates – the famous Uemura relation [19, 20].

Impurities in the $CuO_2$ plane degrade $T_c$ very effectively in cuprates [21 – 23]. This is mainly because of the *d*-wave character of the SC order parameter. It should be noted that isovalent impurities replace the Cu atoms and do not alter the hole content. These impurities break Cooper pairs via unitary scattering [21] and thereby reduce superfluid density, $\rho_s$ [23], and $T_c$. Therefore, both disorder and PG reduces the superfluid density in cuprates. Albeit the mechanisms by which superfluid density is depleted are different for impurity and PG, the primary end effect is the same, a decrease of $T_c$. The effect of in-plane defect was demonstrated quite nicely by Nachumi et al. [24]. It has been shown clearly that $T_c$ varies linearly, in the same fashion, for both hole doping and impurity induced changes in the values of $\rho_s/m^*$, where $\rho_s$ is the superfluid density and $m^*$ is the effective mass of the Cooper pair [24]. Validity of the Uemura relation implies that the dependence of $T_c$ on the superfluid density does not depend significantly on the details of the mechanism via which it is changed. For example, $T_c$ of both Zn doped Y123 and Zn free Y123 (where $\rho_s$ and $T_c$ are changed by varying the oxygen content) fall in the same line when plotted against $\rho_s/m^*$ [24]. These features appear to be quite generic in hole doped copper oxide superconductors. Cuprates belonging to different families show similar qualitative behavior as a function of hole content and in-plane isovalent nonmagnetic disorder [14, 21 – 23].

In this brief report, we have attempted to investigate the effect of in-plane impurity on $T_c$ at a fixed hole content and that due to the *p*-dependent PG on $T_c$ in pure cuprate superconductors. The effects are contrasted and compared. The doping and impurity induced variation of the oxygen isotope effect has been used as a testing ground. We have found strong circumstantial evidence that, as far as $T_c$ degradation is concerned, both in-plane impurities and the PG play qualitatively similar role. Depletion of the superfluid density by both (impurity and the PG) is the common thread that links the experimental observations.

Enhancing the SC transition temperature of cuprates has been a major goal of high-$T_c$ community. The highest $T_c$ achieved so far stands at 164 K in triple $CuO_2$ layer optimally doped Hg-based cuprate under high quasi-hydrostatic pressure [25, 26]. Despite extensive effort of the researchers, $T_c$ of hole doped cuprates has not increased since 1994. The prime reason for this is the inability of the condensed matter community to pin-point the precise mechanism leading to hole pairing in these strongly correlated electronic systems. The optimal structure and composition giving rise to the *right* electronic, phononic, and magnetic excitation spectra for higher $T_c$ remains elusive till now. Circumstantial evidence points towards significant role played by phonons and spin excitation spectrum on the emergence of superconductivity in cuprates [27 – 30]. Several theoretical studies have been carried out within the framework of phonon mediated superconductivity for copper oxide superconductors [28 – 30]. Recent discovery of extremely high critical temperature superconductivity above 200 K in metallic sulfur hydrides under extreme pressure [31] and very recent study of sulfur isotope effect in $H_3S$ [32] have demonstrated clearly the possible role that phonons can play in high-$T_c$ systems. Isotope effect remains the most significant experimental and theoretical tool to explore phonon mediated pairing in SC systems. This provides us with additional motivation to carry out this project.

Rest of the paper is organized as follows: Section 2 describes the framework which is used to analyze the experimental data. Section 3 consists of the discussion on the main findings and the conclusions.

## 2 Proposed framework and analysis of the experimental data

The reduction of superconducting transition temperature due to impurity scattering can be expressed via the Abrikosov-Gorkov (A-G) equation [33] given below.

$$\ln(\frac{T_{c0}}{T_c}) = \kappa[\psi(\frac{1}{2} + \gamma) - \psi(\frac{1}{2})] \qquad (1)$$

Here, $T_{c0}$ is the SC transition temperature of the pure compound, $T_c$ is the transition temperature of the impurity substituted compound, $\Psi$ denotes the di-gamma function, $\kappa$ is a constant of the order of unity which depends on the symmetry of the SC gap function, $\gamma = \Gamma/2\pi k_B T_c$, with $\Gamma$ being the impurity induced pair-breaking scattering rate. For strong (potential) pair-breaking scattering, $\Gamma = x/\pi<N(\varepsilon_F)>$, where $<N(\varepsilon_F)>$ is the thermally averaged electronic energy density of states (EDOS) at the Fermi level ($\varepsilon_F$) and $x$ is the impurity concentration in the $CuO_2$ plane. Interestingly, the Uemura relation implies that $T_c \sim \beta\rho_s$, where $\beta$ is a constant of proportionality (or a weakly varying parameter with hole doping).

This indicates that Eq.1 can be expressed approximately in terms of superfluid density instead of the superconducting transition temperature. It also suggests that as long as the Uemura relation holds, the A-G equation can be used to estimate the fractional change in the superfluid density due to impurity scattering.

The PG correlation in the normal state reduces the low-energy QP spectral weight around the Fermi energy [13, 14, 17, 18, 23]. Even though the exact physics behind the origin and evolution of the PG energy scale, $\varepsilon_g$, with hole content remains a matter of debate, it is largely agreed upon that PG competes with superconductivity [17, 23]. Since PG depletes the EDOS around the Fermi level, QP spectral weight is reduced and the SC pair density is diminished. This conclusion follows directly from the relation $\rho_s \sim \langle N(\varepsilon_F)\rangle \Delta_{sc}$, where $\Delta_{sc}$ is the amplitude of the SC gap. The thermal average is usually taken over an energy window of width ± 2-$3k_B T_c$ centered at the Fermi level. Presence of the PG depletes $\langle N(\varepsilon_F)\rangle$ and therefore, reduces the superfluid density. It should be noted that PG is small near the optimum doping, it increases with underdoping and $T_c$ falls in the underdoped region [7, 13, 17, 18, 23]. An anti-correlation between $T_c$ and $\varepsilon_g$ is seen in all the hole doped cuprates.

Zn is nonmagnetic and isovalent to the in-plane Cu atom that it substitutes. Effect of Zn on $T_c$ has been studied extensively [6, 7, 21 – 23, 34]. We focus on the Zn induced suppression of $T_c$ in $La_{2-x}Sr_xCu_{1-y}Zn_yO_4$ (Zn-LSCO) and $YBa_2(Cu_{1-y}Zn_y)_3O_{7-\delta}$ (Zn-YBCO) superconductors. Fig.1 shows the $T_c$ versus in-plane Zn content data for Zn-LSCO and Zn-YBCO [23, 34 – 37]. It is seen in Fig.1 that the rate of suppression of $T_c$ varies markedly with hole content. Nonmagnetic defects can break Cooper pairs due to the $d$-wave symmetry of the order parameter. The pair-breaking scattering rate is inversely proportional to the thermally averaged EDOS in the vicinity of the Fermi energy. Since this EDOS decreases as the PG energy scale becomes larger with underdoping, $dT_c/dy$ also increases with the decrease of $p$.

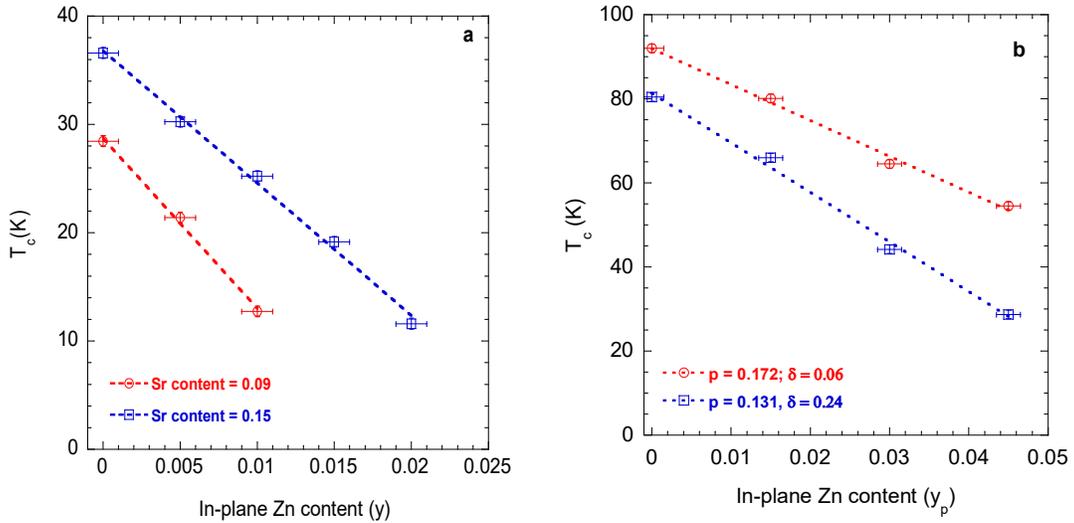

Figure 1. $T_c$ versus in-plane Zn content in (a) $La_{2-x}Sr_xCu_{1-y}Zn_yO_4$ and (b) $YBa_2(Cu_{1-y}Zn_y)_3O_{7-\delta}$ [23, 34 – 37]. The dashed lines are the least-square fits to the data.

The $T_c$ values shown in Fig.1 are accurate within ± 0.5 K. The $p$ values were obtained from the room-temperature thermoelectric power data following Refs. [38, 39] and also from the parabolic $T_c(p)$ relation [5 – 7]. The reported values of $p$ and $\delta$ are accurate up to ± 0.005 and ± 0.02, respectively. For Zn-LSCO the Sr content, $x$, equals the in-plane hole content, $p$. It should be noted that YBCO has two $CuO_2$ planes and a single $CuO_{1-\delta}$ chain per formula unit. Zn atoms reside in the $CuO_2$ planes. Therefore, for this compound, the in-plane Zn content, $y_p = 3y/2$.

Next, we plot $T_c$ versus $T^*$ data in Fig.2. $T^*$ is the characteristic PG temperature ($T^* \equiv \varepsilon_g/k_B$) obtained from the analysis of magnetic bulk susceptibility data [18, 23] and $ab$-plane resistivity data [13, 14]. It is possible to calculate the pair-breaking scattering rate from the values of $T_c$ and $T_{c0}$ using Eq. 1. Both $\Delta T_c$ (= $T_{c0} - T_c$) versus the scattering rate (expressed in energy scale) and $\Delta T_c$ versus $\varepsilon_g$ show approximately linear trend over an extended region in the underdoped side. Here $T_{c0}$ is taken as the maximum $T_c$ at optimum doping for a pure compound. This again provides us with a clue that as far as $T_c$ suppression due to in-plane defect and due to underdoping in defect-free cuprates are concerned, there is possibly a common mechanism at play.

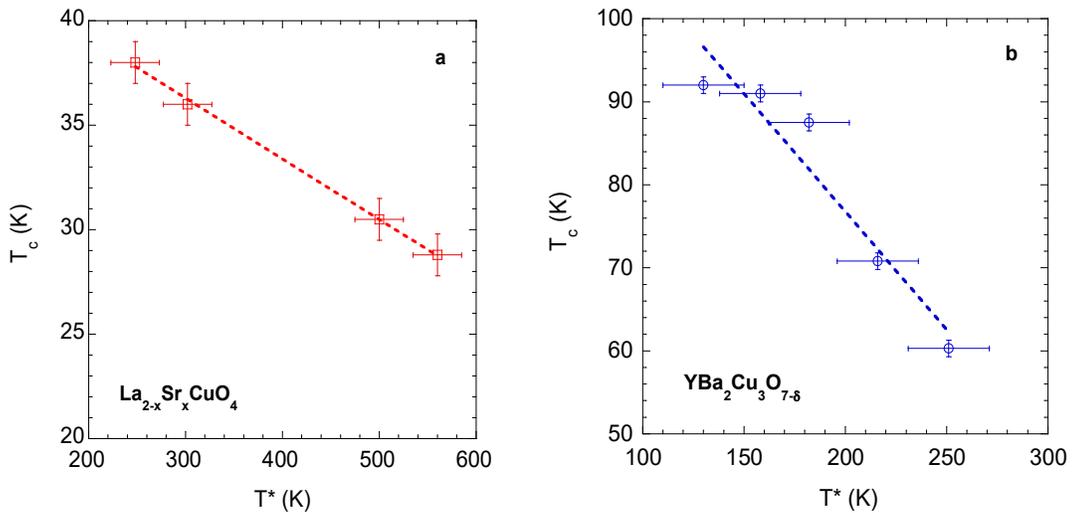

Figure 2. $T_c$ versus $T^*$ for (a) $La_{2-x}Sr_xCu_{1-y}Zn_yO_4$ and (b) $YBa_2(Cu_{1-y}Zn_y)_3O_{7-\delta}$ [6, 13, 14, 18, 23]. The dashed lines are the least-square fits to the data.

Thus far, we have seen that both in-plane impurities at a fixed hole content and the pseudogap as $p$ is varied in impurity free compounds, have same qualitative effect on the superconductive transition temperature. Impurities reduce the superfluid density via pair-breaking and the PG reduces the superfluid density via the removal of low energy QP spectral density. The common Uemura relation for $p$-dependent $T_c$ and that due to Zn substitution lends strong support to the observations presented in this paper. The $p$ and Zn dependent variation of the isotope exponent can also shed significant light on this matter.

In high-$T_c$ cuprates, the isotope exponent (IE), $α$, depends strongly on the number of doped carriers in the CuO$_2$ plane. In general, $α(p)$ rises sharply in the underdoped side and can exceed the canonical BCS value of 0.5 near the so-called 1/8$^{th}$ anomaly [40 – 42]. The value of IE near the optimum doping is very small and remains almost the same (or increases slightly) in the overdoped region [40 – 44]. The problem in interpretation of the $α(p)$ data for cuprates is extremely complicated because of the various electronic correlations present in these remarkable strongly correlated electronic materials. The IE is affected both by spin/charge ordering (the so-called *stripe correlations*) and PG in the QP spectrum [40]. How these correlations are related to Cooper pairing itself is not entirely clear yet [27]. Although, many researchers in the field believes short-range antiferromagnetic fluctuations may give rise to superconductivity, there is notable interest in possible role of lattice on the QP dynamics in cuprates following e.g., isotope dependent ARPES results [45]. A clear understanding of the hole content and disorder dependences of the IE is important to clarify various issues regarding carrier pairing and to reveal the possible interrelations among the underlying electronic correlations present in high-$T_c$ cuprate superconductors.

The IE is defined as $α(T_c) = -(d\ln T_c/d\ln M)$, where M is the isotopic mass. Incorporating this expression for IE with the A-G pair-breaking equation, $α$ can be expressed as a universal function on $T_c$ as follows [44]

$$\frac{α(T_c)}{α(T_{c0})} = \frac{1}{1 - κψ'(\frac{1}{2} + γ)γ} \qquad (2)$$

Here, $ψ'$ is the first derivative of $ψ$. $α(T_{c0}) = -(d\ln T_{c0}/d\ln M)$, IE of the pure compound. Eq. 2 shows that IE of the disordered compound is a function of IE of the pure compound and $T_c$. It also follows that a plot of $T_c/T_{c0}$ versus $α(T_c)/α(T_{c0})$ has a universal character, independent of the details of pair-breaking mechanism or equivalently the physical process that reduces the superfluid density.

We have shown $α(T_c)/α(T_{c0})$ versus $T_c/T_{c0}$ for Zn doped and Zn free LSCO and YBCO [40 - 43] in Fig. 3.

It should be noted that isotope exponents are hard to determine with high degree of precision. It mainly arises from finite superconducting transition width and a subjective element involved in defining the mean-field $T_c$. Both disorder and underdoping tend to increase the superconducting transition width [46]. It is seen in Fig. 3 that the trends of $α(T_c)/α(T_{c0})$ versus $T_c/T_{c0}$ for La$_{2-x}$Sr$_x$Cu$_{1-y}$Zn$_y$O$_4$ and YBa$_2$(Cu$_{1-y}$Zn$_y$)$_3$O$_{7-δ}$ are quite identical within the error bars, irrespective of the physical cause for $T_c$ suppression (underdoping or Zn substitution). The experimental values lie quite close to the universal curve based on the A-G pair-breaking model.

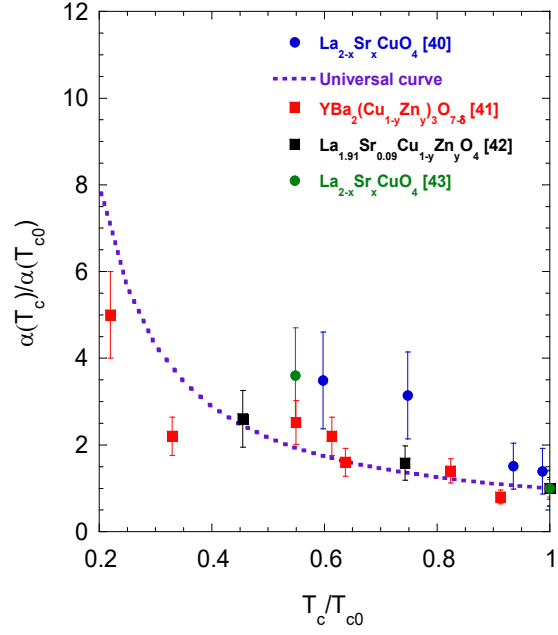

Figure 3. $\alpha(T_c)/\alpha(T_{c0})$ versus $T_c/T_{c0}$ for $La_{2-x}Sr_xCu_{1-y}Zn_yO_4$ and $YBa_2(Cu_{1-y}Zn_y)_3O_{7-\delta}$. The dotted curve represents the universal behavior predicted by Eq. 2 (with $\kappa = 1$).

At this point, it is important to mention that both Eqs. 1 and 2 are derived for conventional BCS superconductors. Cuprates, though strongly unconventional, retains many gross features of BCS superconductivity [47]. The A-G pair-breaking formula is quite general in nature it can be extended easily to *d*-wave superconductivity and for cases where Cooper pairing is mediated by bosons other than phonons. Questions regarding using A-G formalism for high-$T_c$ cuprates have been raised before by Franz et al. [48]. A careful analysis, as presented by Tallon et al. [21], of the mean distance among the in-plane impurities and the in-plane SC coherence length revealed that applicability of A-G formalism in Zn substituted Y123 and La214 remains valid, at least for the impurity concentrations considered in this paper.

## 3 Discussion and conclusions

In this short communications, we have compared the effect of isovalent in-plane impurity substitution and underdoping on the superconducting $T_c$ of hole doped cuprates. While the former breaks Cooper pairs via unitary scattering, the later removes low energy QP sates which otherwise would have been available for SC pairing condensate. It has been proposed here that the primary effect of both in-plane impurities and underdoping is the reduction of superfluid density. The experimental behavior of $\alpha(T_c)/\alpha(T_{c0})$ versus $T_c/T_{c0}$ plot for $La_{2-x}Sr_xCu_{1-y}Zn_yO_4$ and $YBa_2(Cu_{1-y}Zn_y)_3O_{7-\delta}$ compounds and their reasonable agreement with the universal theoretical trend (Fig. 3) lend support to the proposed scenario.

At this point it is worth mentioning that, in a series of studies [49 – 52] Ovchinnikov et al. have proposed a model for normal and superconducting state properties of hole doped cuprates based on inhomogeneity. Ovchinnikov et al. [49 – 52] have suggested that an inhomogeneous distribution of hole content led to local values of superconducting transition temperatures and formation of phase incoherent Cooper pairs at temperatures above the resistive $T_c$ which is due to a percolation threshold. This model invoked doping induced magnetic pair-breaking to describe the $T_c(p)$ behavior and does not seem to differentiate between magnetic impurity substitution and those magnetic defects induced by hole doping. Therefore, the observation reported in this study can be accommodated within this inhomogeneous superconductivity model [49 – 52]. The origin of the PG is an intensely debated issue [6 – 9, 13 – 15, 53, 54]. In the inhomogeneous superconductivity model [49 – 52] the pseudogap arises at high temperatures due to phase incoherent pairing fluctuations. It is important to note that the results shown in this paper involves Zn substituted compounds. Zn is non-magnetic in nature and breaks Cooper pairs via unitary scattering. Nevertheless, the formalism suggested in Refs. 49 – 52, remains relevant due to its general and broad-based nature.

It should be noted that the hole content in the $CuO_2$ plane(s) remains unaltered due to isovalent Zn substitution [38, 39]. The hole content, on the other hand, changes with underdoping. The effective mass of the Cooper pair, $m^*$, may also vary with hole content [55]. These are the factors that have not been taken into account in the simple model put forward in this study. Nevertheless, the similitude of the $\alpha(T_c)/\alpha(T_{c0})$ versus $T_c/T_{c0}$ plots for $La_{2-x}Sr_xCu_{1-y}Zn_yO_4$ and $YBa_2(Cu_{1-y}Zn_y)_3O_{7-\delta}$, with and without in-plane impurities, is indicative that a common cause is at play. Superfluid density is the most likely common thread.